\documentclass[aps,prb,showpacs,amsmath,amssymb,twocolumn,superscriptaddress]{revtex4-1}
\pdfoutput=1

\usepackage{color}
\usepackage{amsmath}
\usepackage{graphicx}
\usepackage{multirow}

\begin{document}

\newcommand{\bn}{{\bf n}}
\newcommand{\bp}{{\bf p}}   
\newcommand{\br}{{\bf r}}
\newcommand{\bk}{{\bf k}}
\newcommand{\bv}{{\bf v}}
\newcommand{\brho}{{\bm{\rho}}}
\newcommand{\bj}{{\bf j}}
\newcommand{\wk}{\omega_{\bf k}}
\newcommand{\nk}{n_{\bf k}}
\newcommand{\eps}{\varepsilon}
\newcommand{\la}{\langle}
\newcommand{\ra}{\rangle}
\newcommand{\be}{\begin{eqnarray}}
\newcommand{\ee}{\end{eqnarray}}
\newcommand{\intl}{\int\limits_{-\infty}^{\infty}}
\newcommand{\dE}{\delta{\cal E}^{ext}}
\newcommand{\SE}{S_{\cal E}^{ext}}
\newcommand{\dsp}{\displaystyle}
\newcommand{\phit}{\varphi_{\tau}}
\newcommand{\p}{\varphi}
\newcommand{\cL}{{\cal L}}
\newcommand{\dphi}{\delta\varphi}
\newcommand{\dbj}{\delta{\bf j}}
\newcommand{\lra}{\leftrightarrow}

\newcommand{\rred}[1]{{#1}}
\newcommand{\skp}[1]{{#1}}
\newcommand{\bblue}{}

\title{Shot noise in the edge states of 2D topological insulators}

\author{P.~P.~Aseev}
\affiliation{Kotelnikov Institute of Radioengineering and Electronics, Mokhovaya 11-7, Moscow, 125009 Russia}

\author{K.~E.~Nagaev}
\affiliation{Kotelnikov Institute of Radioengineering and Electronics, Mokhovaya 11-7, Moscow, 125009 Russia}
\affiliation{Moscow Institute of Physics and Technology, Institutsky per. 9, Dolgoprudny, 141700 Russia}

\date{\today}

\begin{abstract}
We calculate the resistance and shot noise in the edge states of a 2D topological insulator that result
from the exchange of electrons between these states and conducting puddles in the bulk of the insulator.
The two limiting cases where the energy relaxation is either absent or very strong are considered. 
A finite time of spin relaxation in the puddles is introduced phenomenologically. Depending on this time 
and on the strength of coupling between the edge states and the puddles, the Fano factor ${\cal F}=S_I/2eI$ 
ranges  from 0 to 1/3, which is in an agreement with the available experimental data.
\end{abstract}


\maketitle

\section{Introduction}

A principal distinctive feature of 2D topological insulators is the existence of helical edge electronic 
states in which the electron spin projection is locked to the direction of its momentum. For this reason, 
the electrons cannot be backscattered unless the time-reversal symmetry is violated. This topological protection 
of the pair of edge states with opposite spin directions results in the universal value of its conductance
$e^2/h$, which should hold in the absence of spin-flip scattering.\cite{Hasan10} However measurements 
revealed that the conductance appears to be much smaller than this universal value. 
In most cited papers on HgTe/CdTe  quantum wells the suppression of conductance is about 10\%  as the length of the 
conducting channel is about 1 $\mu$m,\cite{Konig07,Roth09} but in some experiments on these systems, the conductance 
decreased by two orders of magnitude and the estimates of the coherence-breaking length for the edge states  were 
much lower.\cite{Gusev11, Grabecki13} The common observation for all the experiments was that the suppression of 
the conductance was weakly temperature-dependent. A similar behavior of the conductance was observed in 
InAs/GaSb/AlSb heterostructures.\cite{Du15,Knez14} These facts had no satisfactory explanation so far 
despite a large number of theoretical papers proposing different mechanisms of electron backscattering in the edge 
states. A number of authors considered spin-flip scattering of electrons on magnetic impurities.  A spin relaxation 
of magnetic impurities due to an interaction with nuclear spins was phenomenologically introduced in 
Ref.~\onlinecite{Lunde12}, and the authors of Ref.~\onlinecite{Altshuler13} assumed that the impurities lack the 
axial symmetry, and therefore the corresponding component of the total spin of the electron 
and the impurity is not conserved.  However this would lead to the Anderson localization of the edge states, 
which is 
not experimentally observed. Some authors also considered mechanisms of inelastic scattering on a point defect  
taking into account a strong electron-electron interaction and  a violation of the  $S_z$ symmetry in the edge states by spin-orbit coupling.\cite{Schmidt12} Apart from the scattering by isolated impurities,  a capture of electrons from the 
edge states into the conducting regions in the bulk of the insulator was also considered  in 
Ref.~\onlinecite{Vayrynen13}, where the formation of these conducting regions was attributed to the potential 
fluctuations because of impurity doping. But even these processes could not explain the weak temperature dependence 
of the resistance. In Ref.~\onlinecite{Essert15}, the authors suggested that the backscattering could result from a 
dephasing of electrons captured from the edge state into a quantum “puddle”  with chaotically arranged scatterers, 
but they could not draw a definite conclusion about the relevance of this mechanism to actual experiments. Two-
particle scattering on a defect with localized spin-orbit coupling in a presence of electron-electron interaction 
was also considered.\cite{Crepin12} This process results in a backscattering of electrons in the edge states and a 
weak temperature dependence of the conductance only if the Luttinger  parameter describing the electron interaction 
in the edge states is very close to the value $K=1/4$. Nevertheless there is a general opinion  that  the most 
probable reason of the observed breaking of topological protection of the edge states is the presence of structural 
defects and inelastic-scattering centers. This opinion is supported by the experiment\cite{Konig13} that revealed 
well-localized scattering centers.

The nonequilibrium electric noise provides an important information about the processes of charge transport, which 
cannot be extracted from measurements of average values. Therefore a comparison of its theoretical value with the 
experimental data could allow one to understand the mechanism of conductance suppression in topological insulators. 
So far, only several theoretical papers on the noise in topological insulators were published, and most of them 
considered the electron tunneling from one edge of the sample to another.\cite{Schmidt11,Rizzo13,Edge15,Dolcini15}  These 
authors neglected the scattering in the edge states themselves and these states were assumed to be noiseless. As far 
as we know, the noise produced by backscattering in these states was calculated only in 
Ref.~\onlinecite{delMaestro13}, where it resulted from the hyperfine interaction of the electrons with nuclear spins 
in a presence of nonuniform spin-orbit coupling. The Fano factor of the calculated noise $S_I/2eI$ appeared to be 
larger  than unity in the large-length limit. In recent experiments on the shot noise  in the edge states of HgTe 
based topological insulators, this ratio varied between 0.1 and 0.3 depending on the sample.\cite{Tikhonov15} This 
suggests that the theoretical model \cite{delMaestro13} is inapplicable to such systems. 

In this paper, we calculate the resistance of the edge states and the nonequilibrium noise in them that result from
the tunnel coupling between the edge states and charge puddles in the bulk of the insulators, which is suggested by
recent experimental results.\cite{Roth09,Konig13,Grabecki13} 
These puddles are believed to form because of inhomogeneous distribution of doping impurities in the adjacent 
layers of material.\cite{Vayrynen13}
We assume that they have a continuous energy
spectrum and the motion of electrons in them is two-dimensional, so that the impurity scattering combined 
with spin-orbit coupling may result in their 
temperature-independent spin relaxation via the Elliott--Yafet\cite{Elliott54} or Overhauser 
mechanism,\cite{Overhauser53} see Ref. \onlinecite{Zutic04} for a review.\cite{1D-no-spin-flip}
Hence the presence of the puddles enables backscattering of 
the electrons in the edge states 
and results in their increased resistance along with a finite shot noise in them. We calculate the noise for the two
limiting cases where the energy relaxation of electrons in the puddles is either absent or very strong. By 
comparing the magnitude of the 
shot noise with the increase in the resistance, one can judge upon the relevance of this model to real topological 
insulators.

The paper is organizes as follows. In Sec. II, we present our model and the kinetic equations for the average 
current and its fluctuations. In Sec. III, we consider the contribution to the resistance and noise in the absence
of energy relaxation in the puddles. In Sec. IV, the opposite limit of a strong energy relaxation is considered,
and Sec. V summarizes the results.

\section{Model and general equations}

Consider a pair of helical edge states with linear dispersion $\eps_p=|p|\,v$ connecting electron reservoirs
that are kept at constant voltages $\pm V/2$. Each of the two directions of the electron momentum is locked to
a definite spin projection, which is labelled by $\sigma=\pm 1$. For simplicity, the interaction between the electrons in these states is neglected. The edge states are tunnel-coupled with electron or hole puddles that 
are formed in the bulk of the insulator because of large-scale potential fluctuations. We also assume that these 
puddles are sufficiently large to have a continuous spectrum and that the  electrons in the puddles are also subject to a spin relaxation because of spin-orbit processes and, in general, to the energy relaxation. 

The distribution functions of electrons in the edge states $f_{\sigma}(x,\eps,t)$ obey the equations
\begin{multline}
 \left(
  \frac{\partial}{\partial t} + {\sigma}v\,\frac{\partial}{\partial x}
 \right) f_{\sigma}(x,\eps,t)
\\ =
 -\sum_i \Gamma_i(x)\,[ f_{\sigma}(x,\eps,t) - F_{i\sigma}(\eps,t)],
 \label{f-eq1}
\end{multline}
where $x$ is the coordinate along the edge of the insulator, $\eps$ is the energy, $\Gamma_i(x)$ is the rate of 
electron tunneling from point $x$ to the puddle $i$,  and $F_{i\sigma}(\eps, t)$  is the spin-dependent 
distribution function of electrons in the puddle $i$. As the conductance of the puddle is much higher than that of the edge states, this distribution functions is  spatially uniform 
inside it and obeys the equation
\begin{multline}
 \frac{\partial F_{i\sigma}}{\partial t} 
 + \frac{1}{2\pi\hbar v \nu_i} \int dx\,
 \Gamma_i(x)\,
 [F_{i\sigma}(\eps,t) - f_{\sigma}(x,\eps,t)] 
\\
 + \frac{1}{2\tau_s}\,(F_{i\sigma} - F_{{i,-\sigma}})
 = I_{\eps}(\eps,t),
 \label{F-eq1}
\end{multline}
where $\nu_i$ is the density of states in puddle $i$, $\tau_s$ is 
the spin-relaxation time, and the collision integral $I_{\eps}$ accounts for the energy 
relaxation but conserves the number of electrons with a given spin projection in the puddle. At the zero 
temperature, the distribution functions of electrons in the right and left reservoirs are Fermi steps, so
the boundary conditions for $f_{\sigma}$ are
\be
\begin{split}
 f_{+}(0,\eps) = 1 - \Theta(\eps-eV/2),
 \\
 f_{-}(L,\eps) = 1 - \Theta(\eps+eV/2),
\end{split}
 \label{f-boundary}
\ee
where $\Theta $ is the Heaviside step function and $V$ is the applied voltage. The current carried by the
edge states is given by 
\be
 I = \frac{e}{2\pi\hbar} \int_{-eV/2}^{eV/2} d\eps\,[f_{+}(x,\eps) - f_{-}(x,\eps)].
 \label{I-eq1}
\ee


In a semiclassical system, the dynamics of fluctuations is conveniently described by a set of Langevin
equations for the relevant distribution functions. These equations are derived by varying the kinetic
equations for the corresponding average quantities and adding Langevin sources to the result of 
variation.\cite{Kogan} The variation of Eq. \eqref{f-eq1} with respect to $f_{\sigma}$ and $F_{i\sigma}$
gives
\begin{multline}
 \left( \frac{\partial}{\partial t} + {\sigma}v\frac{\partial}{\partial x} \right) \delta f_{\sigma}
\\
 = -\sum_i \Gamma_i(x)\,(\delta f_{\sigma} - \delta F_{i\sigma}) + \sum_i \delta J_{i\sigma},
 \label{df-eq1}
\end{multline}
where $\delta J_{i\sigma}(x,\eps,t)$ is the Langevin source related to tunneling of electrons from point 
$x$ of the edge state with spin projection $\sigma$ to puddle $i$ and back. Similarly, the variation of
Eq.~\eqref{F-eq1} gives
\begin{multline}
 \frac{d}{d t}\,\delta F_{i\sigma} 
 + \frac{1}{2\pi\hbar v\nu_i} \int dx\,\Gamma_i\,(\delta F_{i\sigma} - \delta f_{\sigma})
\\
 + \frac{1}{2\tau_s}\,(\delta F_{i\sigma} - \delta F_{i,-\sigma})
\\
 = \delta I_{\eps} 
 - \frac{1}{2\pi\hbar v\nu_i} \int dx\,\delta J_{i\sigma} + \delta{\cal J}_{i\sigma},
 \label{dF-eq1}
\end{multline}
where $\delta{\cal J}_{i\sigma}=-\delta{\cal J}_{i,-\sigma}$ is the Langevin source related to spin-flip 
scattering. The Langevin source
related to energy relaxation is omitted here because it is inessential in the limiting cases considered below.
The Langevin sources $\delta J_{i\sigma}$ and $\delta{\cal J}_{i\sigma}$ may be treated as independent because
they correspond to different scattering processes. As the scattering is assumed to be weak,
it may be considered Poissonian, and the correlation functions of the Langevin sources in these equations may be
written as the sums of outgoing and incoming scattering fluxes.
The spectral density of tunneling-related sources is given by the well-known expression\cite{Blanter00}
\begin{multline}
 \la\delta J_{i\sigma}(x,\eps)\,\delta J_{j\sigma'}(x',\eps')\ra_{\omega}
\\
 =4\pi\hbar v\,\Gamma_i(x)\,
 \delta(x-x')\,\delta(\eps-\eps')\,\delta_{\sigma\sigma'}\,\delta_{ij}\,
\\ \times
 [f_{\sigma}\,(1-F_{i\sigma}) + F_{i\sigma}\,(1-f_{\sigma})],
 \label{dJ2-1}
\end{multline}
while the spectral density of the sources related to the spin-flip scattering in the puddles 
equals\cite{Mishchenko03,Nagaev06}
\begin{multline}
 \la\delta{\cal J}_{i\sigma}(x,\eps)\,\delta{j\cal J}_{\sigma'}(x',\eps')\ra_{\omega}
 = \frac{1}{\tau_s\nu}\,\delta(\eps-\eps')\,\delta_{ij}\,
\\ \times
 (-1)^{(\sigma-\sigma')/2}\,
 \bigl[F_{i\sigma}\,(1-F_{i,-\sigma}) + F_{i,-\sigma}\,(1-F_{i\sigma})\bigr].
 \label{dY2-1}
\end{multline}
The boundary conditions for the fluctuations $\delta f_{\sigma}$  at the ends of the edge states are
\be
 \delta f_{+}(0,\eps) = \delta f_{-}(L,\eps) = 0.
 \label{df-boundary}
\ee
Equations \eqref{df-eq1} - \eqref{dF-eq1} together with the correlation functions \eqref{dJ2-1} -
\eqref{dY2-1} and the boundary conditions \eqref{df-boundary} allow us to calculate the spectral density
of current noise in the edge states.

\section{Purely elastic scattering}

First consider the case where there is no inelastic scattering of electrons in the puddles. Then the collision  
integral in the right-hand sides of Eqs. \eqref{F-eq1} and \eqref{dF-eq1} may be omitted, and the energy 
dependences of the distribution functions are determined solely by the boundary conditions \eqref{df-boundary}. 
Hence at low temperatures the distribution functions $f_{\sigma}$ and $F_{i\sigma}$ have the characteristic 
two-step\cite{Nagaev92} shape: they are equal to 1 at $\eps<-eV/2$, to 0 at $\eps>eV/2$, and to some
position-dependent but energy-independent intermediate value at $-eV/2 <\eps < eV/2$. These partially occupied states give rise
to a finite shot noise that does not vanish even if the conductance is strongly suppressed by the puddles.

\subsection{Scattering off a single puddle}
\label{elasic-single}

\begin{figure}[t]
\includegraphics[width=8.5cm]{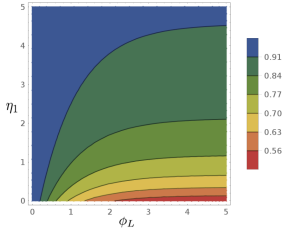}
\caption{(Color online) Contour plot of the conductance  for one puddle given by Eq. \eqref{I-eq2-el-1} 
in coordinates effective 
coupling strength $\phi_L$ -- normalized spin-flip time $\eta_1$.
As the coupling and the spin-flip rate increase, it decreases from  $e^2/2\pi\hbar$ to  
$e^2/4\pi\hbar$}
\label{cond-1}
\end{figure}

First consider the case of a single puddle without energy relaxation, which is described by a single spin-dependent 
electron distribution $F_{\sigma}(\eps)$ and single tunneling rate $\Gamma(x)$. 
In what follows, we will be interested only in the range $-eV/2 <\eps < eV/2$,
where $f_{\sigma}$ and $F_{\sigma}$ are different from 0 and 1  and do not depend on $\eps$. Introducing a new coordinate variable
\be
 \phi(x) = \int_0^x \frac{dx'}{v}\,\Gamma(x'),
 \label{phi}
\ee
one may write the formal solution of stationary Eqs. \eqref{f-eq1} for $-eV/2 <\eps < eV/2$ in the form
\be
\begin{split}
 f_{+}(\phi) =& e^{-\phi} + (1 - e^{-\phi})\,F_{+},
 \\
 f_{-}(\phi) &= (1 - e^{\phi-\phi_L})\,F_{-},
\end{split}
\label{f-eq2-el-1}
\ee
where $\phi_L \equiv\phi(L)$ describes the total strength of the coupling between the puddle and the edge 
states. In terms of the new variable $\phi$, Eq. \eqref{F-eq1} may be recast in the form
\be
 \left( \phi_L + \frac{\pi\hbar\nu}{\tau_s} \right) F_{\sigma}
 - \frac{\pi\hbar\nu}{\tau_s}\,F_{-\sigma}
 = \int_0^{\phi_L} d\phi\, f_{\sigma}(\phi).
 \label{F-eq2-el-1}
\ee
A substitution
of Eq. \eqref{f-eq2-el-1} into Eq. \eqref{F-eq2-el-1} results in a closed system of equation for $F_{\sigma}$ with
the solution
\be
 F_{+} = \frac{1}{2}\,\frac{1 + 2\eta_1}{1 + \eta_1},
 \quad
 F_{-} = \frac{1}{2}\,\frac{1}{1 + \eta_1},
 \label{F-eq3-el-1}
\ee
where $\eta_1 = (1 - e^{-\phi_L})\,\tau_s\,/2\pi\hbar\nu$ is the ratio of the spin-flip time and the dwell time
of an electron in the puddle. 
Substituting these values into Eqs. \eqref{f-eq2-el-1} and making use of the expression for the current
\eqref{I-eq1}, one easily obtains that
\be
 I = \frac{e^2V}{4\pi\hbar}\,
   \frac{1 + e^{-\phi_L} + 2\eta_1} 
        {1 + \eta_1}. 
 \label{I-eq2-el-1}
\ee
Hence the conductance of the system varies from $e^2/2\pi\hbar$ for a weak coupling to the puddle or slow 
spin relaxation in it to the minimal value of $e^2/4\pi\hbar$ for a strong coupling and fast spin relaxation
(see Fig. \ref{cond-1}).

\begin{figure}[t]
\includegraphics[width=8.5cm]{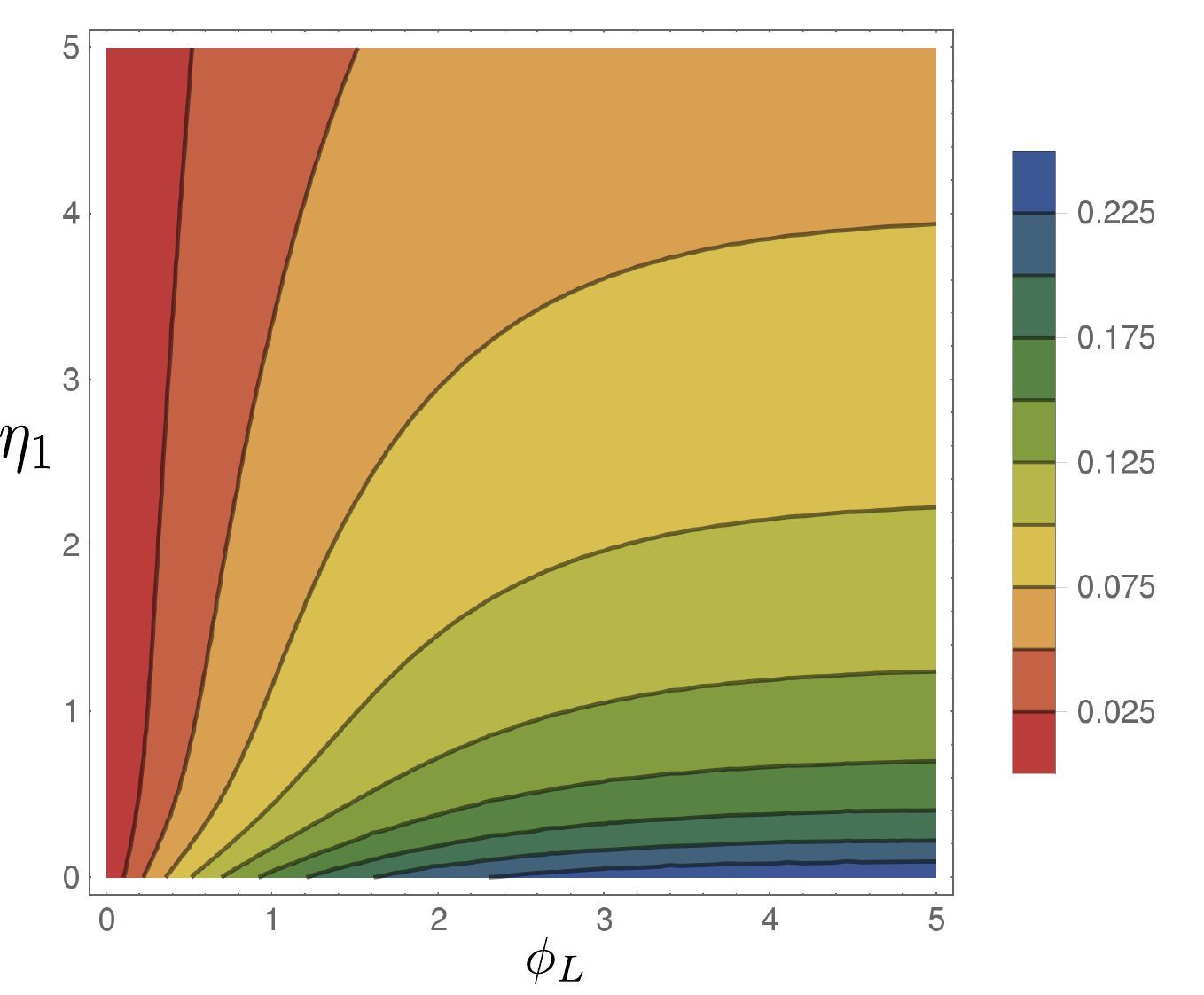}
\caption{(Color online) Contour plot of Fano factor for one puddle in the absence of energy 
relaxation given by Eq. \eqref{Fano-el-1} in coordinates effective coupling strength $\phi_L$ -- normalized spin-flip time $\eta_1$.
The maximum Fano factor corresponds to the maximum resistance of the edge states.}
\label{el-1}
\end{figure}

At low frequencies, the fluctuations of the distribution functions $f_{\sigma}$ are easily obtained from Eq.
\eqref{df-eq1} in the form
\be
\begin{split}
 &\delta f_{+}(\phi) = (1 - e^{-\phi})\,\delta F_{+} 
 + \int\limits_0^{\phi} d\phi'\,e^{\phi'-\phi}\,\Gamma^{-1}\,\delta J_{+},
\\
 &\delta f_{-}(\phi) = (1 - e^{\phi-\phi_L})\,\delta F_{-} 
 + \int\limits_{\phi}^{\phi_L}  d\phi'\,e^{\phi-\phi'}\,\Gamma^{-1}\,\delta J_{+}.
\end{split}
\label{df-eq2-el-1}
\ee
Equation \eqref{dF-eq1} may be rewritten in the quasi-stationary case as%
\begin{multline}
 \left( \phi_L + \frac{\pi\hbar\nu}{\tau_s} \right) \delta F_{\sigma}
 - \frac{\pi\hbar\nu}{\tau_s}\,\delta F_{-\sigma}
\\
 = \int_0^{\phi_L} d\phi\, \bigl(\delta f_{\sigma} - \Gamma^{-1}\,\delta J_{\sigma} \bigr)
 + 2\pi\hbar\nu\,\delta{\cal J}_{\sigma},
 \label{dF-eq2}
\end{multline}
and a substitution of Eqs. \eqref{df-eq2-el-1} results in a closed system of algebraic equations 
for $\delta F_{\sigma}$. Making use again of Eqs. \eqref{df-eq2-el-1} and the linearized Eq. \eqref{I-eq1},
eventually one arrives at the expression for the fluctuation of the current in the form
\begin{multline}
 \delta I = \frac{e}{1 + \eta_1}
 \int d\eps \Bigl[
  \nu\,\eta_1\,\delta{\cal J}_{+}
\\
  +\frac{1}{4\pi\hbar}\int_0^{\phi_L} d\phi\,\Gamma_e^{-1}
   \left( e^{\phi-\phi_L}\,\delta J_{+} - e^{-\phi}\,\delta J_{-}\right)
  \Bigr].
  \label{dI-eq2-el-1}
\end{multline}
Multiplying two instances of Eq. \eqref{dI-eq2-el-1} and making use of the spectral densities of Langevin 
sources Eqs. \eqref{dJ2-1} and \eqref{dY2-1}, one obtains the equation for the spectral density of the noise
\begin{multline}
 S_I = \frac{e^2}{4\pi\hbar}\,\frac{1}{(1 + \eta_1)^2}
 \int d\eps\,  
 \Biggl(
   2\eta_1\,(1-e^{-\phi_L})\,                                                    \\                
   \times                    \bigl[ F_{+}\,(1-F_{-}) + F_{-}\,(1-F_{+}) \bigr] \\
   +\int_0^{\phi_L} d\phi\,
   \Bigl\{
    e^{2(\phi-\phi_L)}\,     \bigl[ f_{+}\,(1-F_{+}) + F_{+}\,(1-f_{+}) \bigr] \\
   +e^{-2\phi}\,             \bigl[ f_{-}\,(1-F_{-}) + F_{-}\,(1-f_{-}) \bigr]
   \Bigr\}
 \Biggr).
 \label{SI-1}
\end{multline}
The Fano factor ${\cal F} = S_I/2eI$ is given by the expression 
\be
 {\cal F} 
 = \frac{1}{4}\,(1-e^{-\phi_L})\,
 \frac{ 1 + e^{-\phi_L} + 4\eta_1 + 4\eta_1^2 + 4\eta_1^3}
      {(1 + e^{-\phi_L} +2\eta_1)(1 + \eta_1)^3}.
 \label{Fano-el-1}
\ee
The contour plot of $\cal F$ is shown in Fig. \ref{el-1}.
It varies from zero for $\phi_L=0$ or infinitely large $\eta_1$ to 1/4 for strong coupling $\phi_L$
and short spin-flip times $\eta_1=0$. Hence the maximum Fano factor corresponds to the maximum resistance
of the edge states. Note that $\cal F$ is not an unique function of the conductance.

\subsection{Multiple puddles in the continuous limit}
\label{elastic-multiple}

\begin{figure}[t]
\includegraphics[width=9cm]{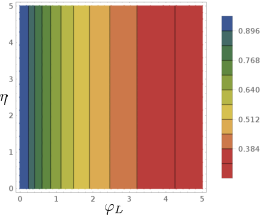}
\caption{(Color online) Contour plot of the conductance  for a continuous distribution of puddles 
given by Eq. \eqref{I-el-c} in coordinates effective coupling strength $\varphi_L$ -- normalized 
spin-flip time $\eta$.
As the coupling and the spin-flip rate increase, it decreases from $e^2/2\pi\hbar$ to 0. }
\label{cond-c}
\end{figure}

Consider now the case where the edge states are weakly tunnel-coupled to many conducting puddles. As the 
distribution functions only slightly change from one puddle to another, it is possible to go to the 
continuum limit and assume that the number $n$ of the puddles per unit length of the insulator edge, the
density of states in the puddles $\nu$, the coupling constant
 $$\Gamma(x) = \frac{1}{\Delta x} \sum_{i \in [x, x+\Delta x]} \int dy\,\Gamma_i(y),$$ 
and the electron distributions
in the puddles are smooth functions of the coordinate $x$. Hence one may just omit the summation over
the puddle number $i$ and replace $F_{i\sigma}(\eps,t)$ by $F_{\sigma}(x,\eps,t)$ in the right-hand side
of Eq. \eqref{f-eq1}. Along with this, one may factor out $f_{\sigma}$ from the integral in the left-hand
side of Eq. \eqref{F-eq1} so that it becomes local in space and 
assumes the form
\be
 \frac{\partial F_{\sigma}}{\partial t} + \frac{1}{\tau_d}\,(F_{\sigma} - f_{\sigma}) 
 + \frac{1}{2\tau_s}\,(F_{\sigma} - F_{-\sigma})
 = 0,
 \label{F-eq1-el-c}
\ee
where $\tau_d(x) = 2\pi\hbar v\,\nu(x)\,n(x)/\Gamma(x)$ is the effective dwell time of an electron in the 
puddle. In the stationary case, Eq. \eqref{F-eq1-el-c} is readily solved for $F_{\sigma}$ giving
\be
 F_{\sigma}(x,\eps) = \frac{1}{2}\,
 \frac{(1 + 2\,\eta)\,f_{\sigma} + f_{-\sigma}}{1 + \eta},
 \label{F-eq2-el-c}
\ee
where $\eta=\tau_s/\tau_d$. A substitution of these values into Eq. \eqref{f-eq1} results in a closed system
of differential equations for $f_{\sigma}$, 
\be
 \sigma v\,\frac{df_{\sigma}}{dx} 
 = -\frac{1}{2}\,\Gamma_e\,\frac{f_{\sigma} - f_{-\sigma}}{1 +\eta}.
 \label{f-eq2-el-c}
\ee
In terms of a new effective coordinate
\be 
 \p(x) = \int_0^x \frac{dx'}{v}\,\frac{\Gamma(x')}{1 + \eta(x')},
 \label{phi-c}
\ee
the solutions of this system may be written as
\be
 f_{+}(\p) = \frac{2 + \p_L - \p}{2 + \p_L},
 \qquad
 f_{-}(\p) = \frac{\p_L - \p}{2 + \p_L},
 \label{f-eq3-el-c}
\ee
where $\p_L \equiv \p(L)$. A substitution of these distribution functions into Eq. \eqref{I-eq1}
gives 
\be
 I = \frac{e^2 V}{\pi\hbar}\,\frac{1}{2 + \p_L},
 \label{I-el-c}
\ee
which suggests that the conductance of the edge states tends to zero as the number of puddles increases
for any finite spin-flip time (see Fig. \ref{cond-c}).

The Langevin equation for the fluctuation $\delta f_{\sigma}(x,\eps,t)$ is obtained from \eqref{df-eq1}
by omitting the subscript $i$ for all the quantities and replacing $\delta F_{i\sigma}$ by 
$\delta F_{\sigma}(x,\eps,t)$. The spectral density of Langevin sources $\delta J_{\sigma}(x,\eps,t)$
is obtained from Eq. \eqref{dJ2-1} in a similar way. 

The Langevin equation for $\delta F_{\sigma}(x,\eps,t)$
becomes local in space and may be written as
\begin{multline}
 \frac{\partial\delta F_{\sigma}}{\partial t} 
 + \frac{1}{\tau_d}\,(\delta F_{\sigma} - \delta f_{\sigma}) 
 + \frac{1}{2\tau_s}\,(\delta F_{\sigma} - \delta F_{-\sigma})
\\
 = -\frac{\delta J_{\sigma}(x,\eps,t)}{2\pi\hbar v n \nu} 
 + \delta{\cal J}_{\sigma}(x,\eps,t),
 \label{dF-eq1-el-c}
\end{multline}
whereas the spectral density of the spin-flip sources takes up the form
\begin{multline}
 \la\delta{\cal J}_{\sigma}(x,\eps)\,\delta{\cal J}_{\sigma}(x',\eps')\ra_{\omega}
 = \frac{1}{\tau_s n \nu}\,\delta(x-x')\,\delta(\eps-\eps')\,
\\ \times
 [F_{\sigma}\,(1-F_{-\sigma}) + F_{-\sigma}\,(1-F_{\sigma})].
 \label{dY^2-el-c}
\end{multline}
The system of equations for $\delta f_{\sigma}$ and $\delta F_{\sigma}$ is solved in a way similar to the
system of kinetic equations for $f_{\sigma}$ and  $F_{\sigma}$, and the fluctuation of current is expressed in terms of the Langevin sources as
\begin{multline}
 \delta I 
 = \frac{e}{2\pi\hbar}  \int d\eps \int_0^{\p_L} \frac{d\p}{2 + \p_L}\,
\\ \times
 \bigl[(\delta J_{+} - \delta J_{-})/\Gamma + 2\,\tau_s\, \delta{\cal J}_{+}\bigr].
 \label{dI-el-c}
\end{multline}
Hence the expression for the spectral density of current fluctuations is of the form
\begin{multline}
 S_I = \frac{e^2}{\pi\hbar}\,\frac{1}{(2 + \p_L)^2} 
 \int d\eps \int_0^{\p_L}\,\frac{d\p}{1 + \eta}\,
 \Bigl\{
  f_{+}\,(1 - F_{+}) 
\\
  + F_{+}\,(1 - f_{+}) + f_{-}\,(1 - F_{-}) + F_{-}\,(1 - f_{-})
\\
  + 2\,\eta\,\bigl[ F_{-}\,(1 - F_{+}) + F_{+}\,(1 - F_{-}) \bigr]
 \Bigr\}.
\label{SI-c}
\end{multline}
\begin{figure}[t]
\includegraphics[width=8.5cm]{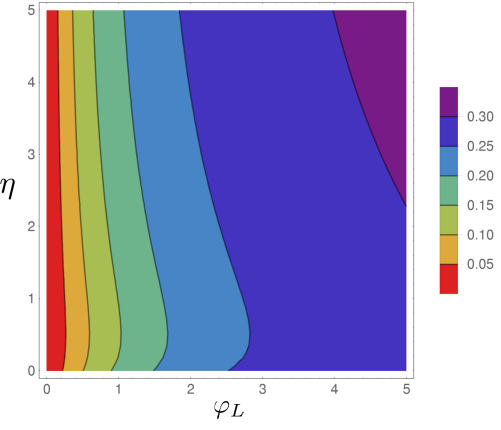}
\caption{(Color online) Contour plot of Fano factor for a continuous distribution of puddles in the 
absence of energy relaxation in coordinates effective coupling strength $\p_L$ -- normalized spin-flip 
time $\eta$ according to Eq. \eqref{Fano-el-c}. The Fano factor vanishes in the ballistic regime and tends to its maximum value 1/3 regardless 
of $\eta$ if the conductance of the edge states tends to zero.}
\label{el-c}
\end{figure} 
The calculation can be carried to the end only if the spatial dependence of $\eta$ is
specified. For simplicity, assume that it is constant. Together with Eqs. \eqref{f-eq3-el-c} 
and \eqref{F-eq2-el-c}, it results in a Fano factor
\begin{multline}
 {\cal F} = 
 \frac{\p_L}{3}\,
 \Bigl[  \p_L\,(\p_L+6)(1+\eta)^3
\\
        + 12\,\eta\,(\eta^2 + \eta + 1) + 6
 \Bigr]
 \left/\Bigl[\right.
    (2+\p_L)(1+\eta)
 \Bigr]^3.
 \label{Fano-el-c}
\end{multline}
The contour plot of Eq. \eqref{Fano-el-c} is shown in Fig. \ref{el-c}.
The Fano factor vanishes in the ballistic limit and tends to its maximum value 1/3 regardless of 
$\eta$ if the conductance
of the edge states tends to zero. This behaviour is reminiscent of multimode  diffusive 
wires.\cite{Nagaev92,Beenakker92}

\section{Strong energy relaxation}

Consider now the opposite case of strong energy relaxation in the puddles. At zero temperature, the
distribution functions of electrons in the puddles have a step-like shape 
$F_{i\sigma}(\eps,t)= \Theta(\mu_{i\sigma} -\eps)$, where $\mu_{i\sigma}(t)$ is the spin-
dependent chemical potential of electrons in puddle $i$. However despite the strong energy relaxation, the shot
noise in such a system is still possible because in general $\mu_{i+} \ne \mu_{i-}$ and there is a
spin imbalance in the puddles. 

It is convenient to introduce the excess densities of electrons in the edge states
\be
 \rho_{\sigma}(x,t) 
 = \int\frac{d\eps}{2\pi\hbar v}\,\bigl[f_{\sigma}(x,\eps,t) - \Theta(-\eps)\bigr].
 \label{rho-in}
\ee
Integrating Eq. \eqref{f-eq1} over the energy results in an equations for $\rho_{i\sigma}(x,t)$ 
of the form
\be
 \left(
  \frac{\partial}{\partial t} + \sigma v\,\frac{\partial}{\partial x}
 \right) \rho_{\sigma}
 =
 -\sum_i \Gamma_i\,
 \left( \rho_{\sigma} - \frac{\mu_{i\sigma}}{2\pi\hbar v}\right),
 \label{rho-eq1}
\ee
which should be supplemented by the boundary conditions
\be
 \rho_{+}(0) = \frac{eV}{4\pi\hbar v},
 \qquad
 \rho_{-}(L) = -\frac{eV}{4\pi\hbar v}.
 \label{rho-boundary}
\ee
Upon integrating Eq. \eqref{dF-eq1} over the energy, the inelastic collision integral drops out
because the inelastic scattering conserves the total number of particles,
and one arrives at the equation 
\begin{multline}
 \frac{\partial \mu_{i\sigma}}{\partial t} 
 + \frac{1}{2\pi\hbar v \nu_i} \int dx\,
 \Gamma_i(x)\,
 (\mu_{i\sigma} - 2\pi\hbar\rho_{\sigma})
\\
 + \frac{1}{2\tau_s}\,(\mu_{i\sigma} - \mu_{i,-\sigma})
 = 0.
 \label{mu-eq1}
\end{multline}
Equations \eqref{rho-eq1} and \eqref{mu-eq1} together with the boundary conditions \eqref{rho-boundary}
form a complete system for determining $\rho_{\sigma}$ and $\mu_{i\sigma}$, and the current flowing
through the edge states equals $I = ev\,[\rho_{+}(x) - \rho_{-}(x)]$.

As the coefficients in Eqs. \eqref{f-eq1} and \eqref{F-eq1} are assumed to be energy-independent, the
energy relaxation in the puddles does not affect the average current. Moreover,
$\rho_{\sigma}(x)$ and $\mu_{i\sigma}$ may be obtained just by integrating $f_{\sigma}$ and $F_{i\sigma}$
obtained for the elastic case over the energy. Things are different if the spectral density of noise is 
considered because the correlation functions \eqref{dJ2-1} and \eqref{dY2-1} are bilinear functions of 
$f_{\sigma}$ and $F_{i\sigma}$. Though the expressions for the spectral density of current noise in terms
of the average distribution functions remain the same, the resulting values appear to be different.

\subsection{Single puddle}

As in the fully elastic case, we start by considering a system with only one puddle. The average current 
in it is given by Eq. \eqref{I-eq2-el-1}. The spin-dependent
chemical potentials of electrons in the puddle may be obtained either by solving Eqs. \eqref{rho-eq1} and 
\eqref{mu-eq1} or by making use of the elastic distribution function \eqref{F-eq3-el-1} and integrating 
the difference $F_{\sigma}(x) - \Theta(-\eps)$ over the energy. This gives us
\be
 \mu_{\sigma} = \sigma\,\frac{eV}{2}\,
 \frac{\eta_1}{1 + \eta_1}.
 \label{mu-in-1}
\ee
The distribution functions are easily obtained by solving Eq. \eqref{f-eq1} and equal
\be
 f_{+}(\phi,\eps) =
 \begin{cases}
  0,         &         \eps>eV/2  \\
  e^{-\phi}, & \mu_{+}<\eps<eV/2  \\
  1,         &         \eps<\mu_{+}
 \end{cases}
 \label{f+in-1}
\ee
and
\be
 f_{-}(\phi,\eps) =
 1 - f_{+}(\phi_L-\phi,-\eps)
 \label{f-vsf+}
\ee
because of the electron -- hole symmetry.
The fluctuation of the current and the spectral density of its noise are given by Eqs.
\eqref{dI-eq2-el-1} and \eqref{SI-1}. The substitution of $f_{\sigma}$  from Eqs. \eqref{f+in-1} and
\eqref{f-vsf+} and $F_{\sigma}(\eps) = \theta(\mu_{\sigma} - \eps)$ with $\mu_{\sigma}$ from Eq.
\eqref{mu-in-1} into Eq. \eqref{SI-1} results in a Fano factor
\begin{figure}[t]
\includegraphics[width=8.5cm]{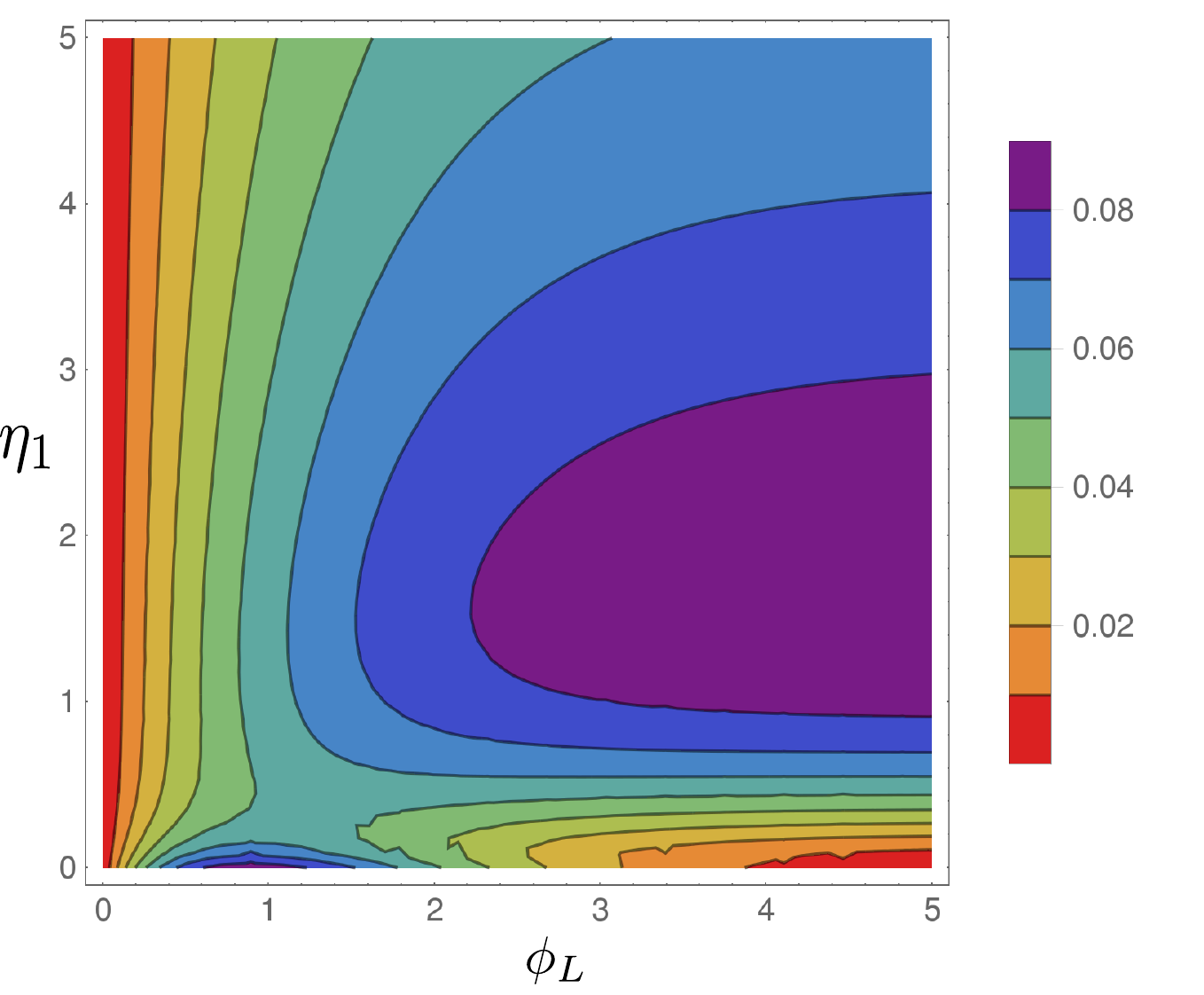}
\caption{(Color online) Contour plot of Fano factor for a single puddle with a strong energy 
relaxation in coordinates effective coupling strength $\phi_L$ -- normalized spin-flip time $\eta_1$
according to Eq. \eqref{Fano-in-1}.
The maximum of Fano factor at $\eta_1=0$ results from the randomness of tunneling between the puddle and
the edge states. The maximum at $\eta_1\approx 1.6$ stems from the randomness of spin-flip scattering
in the puddle.}
\label{in-1}
\end{figure}
\be
 {\cal F} = \frac{1}{2}\,
 \frac{ 1 - e^{-\phi_L} }{ 1 + e^{-\phi_L} + 2\,\eta_1 }\,
 \frac{ e^{-\phi_L} + 2\,\eta_1^2 }{ (1 + \eta_1)^2 }.
 \label{Fano-in-1}
\ee
The contour plot of the Fano factor is shown in Fig. \ref{in-1}. It exhibits a more complicated behaviour then in the
elastic case and shows two separate maxima. One of them ${\cal F} \approx 0.086$ corresponds to the limit of fast 
spin relaxation $\eta_1=0$ and $\phi_L \approx 1$ and results from random tunneling between the puddle and the 
edge states. The other maximum is nearly of the same magnitude ${\cal F} \approx 0.09$ and corresponds to the 
limit of strong puddle -- edge state coupling $\phi_L \to \infty$ and moderate spin relaxation 
$\eta_1 \approx 1.6$. It stems from
random spin-flip scattering in the puddle. 

\subsection{Continuous limit}

\begin{figure}[t]
\includegraphics[width=8.0cm]{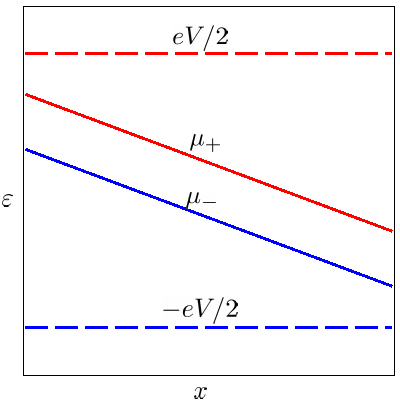}
\caption{The approximate coordinate dependence of the chemical potentials for spin-up and spin-down
electrons in the puddles for the strong energy relaxation. There are finite jumps between the spin-dependent
chemical potentials of the left and right reservoirs and potentials of the puddles.}
\label{mu-fig}
\end{figure}

If there are many puddles weakly coupled to the edge states and the energy relaxation in the puddles is strong, 
one may consider the continuous limit much like as in Section \ref{elastic-multiple}. To this end, we introduce
the coordinate-dependent distribution function of electrons in the puddles $F_{\sigma}(x,\eps,t)= 
\Theta(\mu_{\sigma} -\eps)$, where $\mu_{\sigma}(x,t)$ is the local spin-dependent chemical potential of electrons 
in the puddle at point $x$. The excess densities of electrons in the edge states \eqref{rho-in} obey Eq.
\eqref{rho-eq1} with $\mu_{i\sigma}$ replaced by $\mu_{\sigma}(x)$, and Eq. \eqref{mu-eq1} takes up the form
\be
 \frac{\partial\mu_{\sigma}}{\partial t}
 + \frac{1}{\tau_d}\,(\mu_{\sigma} -2\pi\hbar v\rho_{\sigma})
 + \frac{1}{2\tau_s}\,(\mu_{\sigma} - \mu_{-\sigma}) =0.
 \label{mu-eq1-in-c}
\ee
The solution of these equations is easily obtained,  and in terms of the variable $\p$  \eqref{phi-c},
the spin-dependent chemical potentials may be presented in the form
\be
 \mu_{\sigma}(\p) = \frac{eV}{2}\,
 \frac{ (1+\eta)(\p_L-2\,\p)  + 2\,\sigma\eta }{ (2+\p_L)(1+\eta) }.
 \label{mu-in-c}
\ee
An approximate coordinate dependence of the potentials is shown in Fig. \ref{mu-fig}.
Note that there is a finite jump between the chemical potentials of the reservoir and the puddles at the 
left end 
\be 
 \Delta\mu = \frac{eV}{2} - \mu_{+}(0) = \frac{eV}{ (1+\eta)(2+\p_L) }
 \label{D_mu}
\ee
and a similar jump at the right end.

The current is given by the same Eq. \eqref{I-el-c} as in the purely elastic case. The equation for
the fluctuation of current  and the expression for its spectral density are the same as
Eqs. \eqref{dI-el-c} and \eqref{SI-c}, but the distribution functions $f_{\sigma}$ and $F_{\sigma}$ are now
different. To calculate $f_{\sigma}$ explicitly, one has to specify the coordinate dependence of $\eta$, and we 
assume it to be constant like in Section \ref{elastic-multiple}. Solving Eq. \eqref{f-eq1} readily gives
\be
 f_{+}=
 \begin{cases}
 1,  & \eps<\mu_{+}(\p)
 \\
 \exp\!\left[ \frac{\mu_{+}(\p)-\eps}{\Delta\mu} \right],   &
 \mu_{+}(\p)<\eps<\mu_{+}(0)
 \\{} & {}\\
 \exp\!\left[ \frac{\mu_{+}(\p)-\mu_{+}(0)}{\Delta\mu} \right], &
 \mu_{+}(0)<\eps<eV/2
 \\
 0,  & \eps>eV/2.
 \end{cases}
 \label{f+in-c}
\ee 
and $f_{-}$ is related to $f_{+}$ by the electron--hole symmetry condition Eq. \eqref{f-vsf+}.
\begin{figure}[t]
\includegraphics[width=8.5cm]{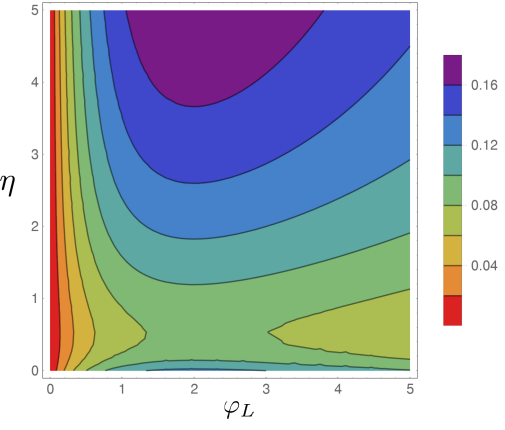}
\caption{(Color online) Contour plot of Fano factor for a continuous distribution of puddles for a strong 
energy relaxation given by Eq. \eqref{Fano-in-c} in coordinates effective coupling strength $\p_L$ -- 
normalized spin-flip time $\eta$.
The smaller maximum of $\cal F$ is located at $\eta=0$, and the larger is reached at $\eta\to\infty$. 
$\cal F$ varies from 0 to 1/4.}
\label{in-c}
\end{figure}
The resulting Fano factor equals
\be
 {\cal F} = 
 \frac{\p_L}{ (2+\p_L)^2 }\,
 \frac{1 + 2\,\eta^2}{ (1 + \eta)^2 }.
 \label{Fano-in-c}
\ee
The contour plot of this equation is shown in Fig. \ref{in-c}. 
Much like as in the case of a single puddle, the Fano factor exhibits two isolated maxima. Both of them 
correspond to $\phi_L=2$, i.~e. to the conductance $e^2/4\pi\hbar$. The smaller maximum ${\cal F} = 1/8$ 
is located at $\eta=0$, and the larger maximum ${\cal F} = 1/4$ is reached at $\eta\to\infty$. Hence 
$\cal F$ varies from 0 to 1/4, but unlike in the elastic case, it vanishes in the limit of zero conductance 
regardless of $\tau_s$.

\section{Conclusion}

In conclusion, we have calculated the conductance and shot noise of a pair of edge states in a 2D
topological insulator using a semi-phenomenological model of conducting puddles in the bulk of material
that can exchange electrons with the edge states. We have considered two versions of this model. The first
version involves one puddle with arbitrary coupling to the edge states, and the second version involved
a continuum of puddles weakly coupled by tunneling to these states. The rate of spin relaxation in the puddles 
was assumed to be finite, and the energy relaxation in them was assumed to be either absent at all or very fast.

In the case of a single puddle without energy relaxation, the conductance decreases with increasing coupling and spin-relaxation rate from $e^2/2\pi\hbar$ to $e^2/4\pi\hbar$. Along with this, the Fano factor
increases from 0 to 1/4.

In the continuum limit without energy relaxation, the conductance tends to zero as the coupling and 
spin-relaxation rate increase, while the Fano factor increases from 0 to 1/3, as in diffusive metals. 
One may think that in the
most realistic case of several puddles strongly coupled to the edge states, $\cal F$ lies somewhere 
between 1/4 and 1/3.

The presence of a strong energy relaxation does not change the conductance but significantly changes the noise. 
The maximum values of the Fano factor are lower than in the elastic case and are now reached at intermediate 
values of conductance. Moreover, $\cal F$ is a nonmonotonic function of both coupling and spin-flip rate and
vanishes in the limit of zero conductance.

The experimental values of the Fano factor for the edge states in HgTe topological insulators\cite{Tikhonov15} 
vary between 
0.1 and 0.3, which roughly agrees with the above model of the noise. However to reliably 
distinguish between different versions of this model, one has to carefully correlate the Fano factor 
of the sample with 
its conductance, which has yet to be done.

\begin{acknowledgments}
We are grateful to V. S. Khrapai for a useful discussion. This work was supported by Russian Science Foundation 
under the grant No. 16-12-10335.
\end{acknowledgments}


\begin{thebibliography}{99}
\bibitem{1D-no-spin-flip}
These mechanisms of spin-flip scattering are inefficient in strictly one-dimensional
systems because the Hamiltonian of spin-orbit interaction involves the vector product of electron
momentum and the potential gradient,\cite{Zutic04} which is zero if they are parallel or antiparallel.
\end{thebibliography}
\end{document}